\documentclass[12pt,preprint]{aastex}

\slugcomment{version of 03-06-04; printed on \today}

\shorttitle{ Arcs in a flat lambda-dominated universe }

\shortauthors{Wambsganss, Bode, \& Ostriker }

\begin{document}

\title{
	Giant Arc Statistics 
	In Concordance With A Concordance LCDM Universe}

\author{Joachim Wambsganss$^1$, Paul Bode$^2$, and 
		Jeremiah P.~Ostriker$^{3,2}$}

\affil{$^1$ Institut f\"ur Physik, Universit\"at Potsdam, 
	14467 Potsdam, Germany}
\affil{$^2$ Dept. of Astrophysical Sciences, Princeton University,
    Princeton, NJ 08544}
\affil{$^3$ Institute of Astronomy, Cambridge University, 
	Madingley Road, Cambridge, UK}
\email{jkw@astro.physik.uni-potsdam.de,bode@astro.princeton.edu,
		jpo@ast.cam.ac.uk}

\begin{abstract}

The frequency of giant arcs--- 
highly distorted and strongly gravitationally 
lensed background galaxies---
is a powerful test for cosmological models. 
Previous comparisons of arc statistics for the currently
favored concordance cosmological model
(a flat lambda-dominated universe)
with observations have shown an apparently large discrepancy.
We present here new ray-shooting results, based
on a high resolution ($1024^3$ particles in a $320 h^{-1}$Mpc box)
large-scale structure simulation
normalized to the WMAP observations.
We follow 
light rays through a pseudo-3D matter distribution approximated
by up to 38 lens planes,  
and evaluate the occurrence of arcs for various source redshifts.
We find that the frequency of strongly lensed background sources
is a steep function of source redshift: 
the optical depth for giant arcs increases by a factor of five 
when the background sources are moved from redshift $z_s=1.0$
to $z_s=1.5$.
This is a consequence of a moderate decrease of the critical
surface mass density for lensing, combined with the
very steep cluster mass function at the high mass end.
Our results are consistent with those 
of Bartelmann et al. (1998)
if we--- as they did---
restrict all sources to be exactly at $z_s=1$. But if we
allow  for a more realistic distribution of source redshifts
extending to or beyond $z_s \ge 1.5$, 
the apparent discrepancy vanishes: the frequency of arcs
is increased by about a factor of ten as compared to previous
estimates, and results in roughly one arc per 20 square degrees
over the sky. 
This prediction for an LCDM model is then in good agreement with
the observed frequency of arcs.
Hence we consider the ``missing arc''  problem for
a concordance  LCDM cosmology to be solved.

\end{abstract}


\section{Introduction}

The occurrence of giant luminous arcs as a result of
strong gravitational lensing by galaxy clusters
is a potentially strong test for cosmological models.  
With the recent WMAP results (Spergel et al. 2003), 
parameters of viable models are now fixed at 
very high redshift, in addition to the  well-known
constraints at redshift zero.
Strong lensing is one of the very few tools
which allows us to probe the intermediate redshift regime,
redshifts $0.3 \le z \le 5.0$, 
in order to confirm the
cosmic background radiation (CBR) results and to break 
the strong degeneracies permitted by even the best
CBR results 
(cf. Bridle et al. 2003).

In recent years, 
this strong lensing test---
formulated as the frequency of
giant arcs---  has been one of the few 
astrophysical results which seemed to disfavor
a flat lambda-dominated universe (LCDM). 
Bartelmann et al. (1998) (hereafter B98) had shown that an LCDM
model appears to underpredict the observed occurrence of arcs
by a factor of 5 to 10.
By simulating the cluster population as singular
isothermal spheres distributed according to Press-Schechter theory,
Cooray (1999a, 1999b)  had found 
no discrepancy between predicted and
observed number of arcs for a flat lambda-dominated universe.
Several attempts have been undertaken to 
determine the reason for the apparent disagreement between
the observed number of arcs and the predictions according to B98:
Meneghetti et al. (2000) investigated whether 
including individual galaxies in the dark matter simulations
of galaxy clusters could change this, but
found only a small effect. 
Similarly, Flores, Maller \& Primack (2000)
found that the galaxy contribution can enhance the cross section
by up to 15\%--- not enough to explain the discrepancy.
Meneghetti, Moscardini \& Bartelmann (2003) 
looked into the question of whether the contribution of the cD
galaxy to the cluster lensing can increase the cross section
for arcs significantly
and concluded that 
this can only moderately enhance the probability
for the production of arcs and cannot explain the discrepancy
between predicted and observed arcs.
Zaritsky \& Gonzalez (2003) re-investigated 
the observational side 
of the arc statistics by 
studying the frequency of strong lensing in a subsample of the
Las Campanas Distant Cluster Survey  (LCDCS). 
Their results (though based on a small number of cases)
confirm earlier observational results,
and also
agree with previous arc studies based on X-ray selected clusters
(Gioia \& Luppino 1994).
Similarly, Gladders et al. (2003) find that the Red-Sequence
Cluster Survey (RCS) has too many arcs compared to the expectation
from B98.

We present here a new analysis of the predicted frequency of arcs
produced by strong lensing of galaxy clusters in a
concordance LCDM model consistent with CMB and low redshift
observations.
We follow light rays through a pseudo 
three-dimensional matter distribution,  drawn from a state-of-the-art
high resolution large-scale structure simulation. 
Our analysis  shows that the expected number of giant arcs in
an LCDM universe is significantly higher than previously thought,  
and is in fact in good agreement with the observations.
In the next section we briefly describe our simulations. In Section 3
we present the results in terms of arc statistics. In the 
discussion (Section 4) we compare our simulations and results
with those of B98 and explain the 
differing conclusions.

\section{Simulations}

In order to simulate the three-dimensional lensing effect 
of a universe filled with dark matter, we first
performed  a large-scale structure simulation with a 
Tree-Particle-Mesh (TPM) code (Bode \& Ostriker 2003). 
TPM uses the Particle-Mesh method for long-range forces and a 
tree code for sub-grid resolution; individual
isolated, overdense regions are 
each treated as a separate tree, thus ensuring efficient 
use of parallel computers.  We used the following
parameters for our cosmological model: 
matter content   $\Omega_{\mathrm M}=0.3$, 
cosmological constant $\Omega_\Lambda=0.7$, 
Hubble constant $H_0=70$ km/sec/Mpc, 
linear amplitude of mass fluctuations $\sigma_8$=0.95,
and primordial power spectral index $n_s$=1.
These parameters are consistent with the 1$\sigma$ WMAP derived
cosmological parameters (Spergel et al. 2003, Table 2).
The simulations were performed in 
a box with a comoving side length of  
$L=320 h^{-1}$Mpc.
We used $N=1024^3=1,073,741,824$ particles,
so the individual particle mass is
$m_{\rm p}=2.54\times 10^9 h^{-1}$ M$_\odot$.
The cubic spline softening length was set to $\epsilon=3.2 h^{-1}$ kpc,
producing a ratio of    box size to softening length of $L/\epsilon =10^5$.
The output was stored at 19 redshift values out to 
$z \approx 6.4$, such that the centers of the saved
boxes matched comoving distances of 
$(160 + n \times 320) h^{-1}$Mpc, where $n=0,...,18$. 

To produce lens planes, first the box was divided
into $9\times 9$ separate square cylinders running the
length of the box.  This was done for
three orthogonal projections, leading to 243 subvolumes.
Two lens planes were produced for each subvolume by
bisecting along the line-of-sight and projecting the mass
in each $160 h^{-1}$Mpc--long volume onto a plane.
At the highest redshift, each plane has a side length of
$35.6 h^{-1}$Mpc and contains $800^2$ pixels, making
the pixel size $44.4 h^{-1}$kpc comoving.  At lower redshifts,
we kept the number of pixels constant but decreased the
size of the planes, such that the opening angle of about
20 arcmin (set by the sidelength of the highest redshift source plane)
remains constant.  Thus in the lowest redshift box the lensing
planes are $1.9 h^{-1}$Mpc on a side and the pixel size is $2.3
h^{-1}$kpc.  Because at lower redshifts the plane does not cover 
all of the corresponding subvolume, there is a random offset
perpendicular to the line of sight within each subvolume.

Light rays are then propagated backwards through these lens planes,
beginning with a regular $960^2$ grid at the lowest redshift lens plane
(i.e. the image plane) and working to higher redshift
(cf. Wambsganss et al. 1995, 1998).
For each realization of our ray shooting simulations,
one of the 243 pairs of lens planes was taken randomly from each 
redshift; thus the chance of repeating structures is very small.
Altogether we performed 100 different realizations.
For each of these, the lensing  properties are evaluated at seven
different source redshifts: 
$z_s=$ 0.5, 1.0, 1.5, 2.5, 3.7, 4.8, and 7.5.
More details on the simulations can be found in Wambsganss,
Bode \& Ostriker (2003).

\section{Arc Statistics}

To analyze arc statistics, it is first necessary to 
identify those strongly lensed background sources to be 
interpreted as arcs. 
For a given realization and a specific  source redshift, we populated
the source plane by a regular grid of 800 by 800 sources, with
separations of about 1.5 arcseconds. 
For each of these sources, 
we determined the image position(s) and magnification(s) 
in the image plane.
We then selected only multiple images and demanded that at
least one of the images had to have a magnification higher
than a certain ``threshold'' magnification. We chose
five different values for this threshold
magnification: $\mu_{\mathrm{arc}} \ge$  5, 10, 15, 20, and 25.

A strongly lensed source is almost always highly distorted in
the tangential direction, i.e. 
for circular sources, the magnification 
$\mu_{\mathrm{arc}}$
is very similar
to the length-to-width ratio  $r_{\mathrm{arc}}$
(this is exactly true for isothermal spheres).
In general, however, 
the magnification of an image can be {\em  both}
in the tangential and radial direction. 
So, in principle, there
could also be highly magnified {\em undistorted} images
(Williams \& Lewis 1998).
However,   as shown by Williams \& Lewis (1998),
for massive clusters with realistic profiles,
almost all highly magnified images are 
indeed strongly elongated distorted  arcs. 
We confirmed this by randomly checking various highly 
magnified images in our simulations; this has been
found by other groups simulating lensing by
galaxy clusters as well (Meneghetti 2003, personal communication).
Hence, from here on we will take the length-to-width ratio
of highly magnified images to be equal with their magnification: 
$r_{\mathrm{arc}}=\mathrm{length/width} \approx \mu_{\mathrm{arc}}$.

In order to evaluate the occurrence of arcs for the LCDM concordance
model, we 
determine for the seven chosen source redshifts
the frequency of multiply imaged sources 
containing at least one image with magnification
greater than a given threshold $\mu_{\mathrm{arc}}$.
Figure 1 shows, as a function of source redshift,
the distribution of this lensing optical depth
for arcs with magnifications
$\mu_{\mathrm{arc}} \ge$  5, 10, 15, 20, and 25.
As can be seen, 
the probability for strong (arc) lensing is 
a very steep function of $z_s$ 
for source redshifts between 
$0.5 \le z_s \le 2.5$;  the distribution
gets somewhat shallower for sources at 
$z_s \ge 2.5$. This is independent of the chosen
threshold magnification (or length-to-width ratio).

In order to compare our results with those
of B98,
we need to  look at a source redshift of $z_s=1.0$ and
a length-to-width ratio of $r \ge 10$: 
our value for this particular optical depth 
(red squares in Figure 1; see also Table 1) is
$ p(r \ge 10, z_s=1.0) = 3.8 \times 10^{-7}$, 
about 12\%
higher than the B98 value,
indicating relatively good agreement between our 
result and the B98 value for 
this particular source redshift. 
The slightly higher result here
can be partly understood as a consequence of our higher mass and
spatial resolution (by factors of about five), 
and the fact that we consider a three-dimensional matter
distribution.
Our normalization (proportional to $\sigma_8 \Omega^{0.6}$) 
may be slightly higher than that adopted by B98.
Using the method of Bode et al. (2001) we estimate the mass 
(within an Abell radius) at
which the cumulative number density of clusters equals
$2\times 10^{6} h^3$Mpc$^{-3}$ to be $5\times 10^{14} h^{-1}M_\odot$;
some of the clusters used by B98 may have been less massive than this.
Given the differences in technique, the agreement between the two
numerical results for sources at $z_s=1$ is remarkably good.

Our analysis of the occurrence of arcs as presented here
is ``conservative'' in the following sense: 
the underlying N-body simulations use only dark matter 
particles, i.e. there is no baryonic component included. As
a consequence, 
the density profiles of the inner cores of the galaxy clusters 
are probably too shallow. 
Inclusion of baryons will lead 
to a steepening of the surface mass density profiles,  which
in turn would produce additional giant arcs 
(cf. Li and Ostriker 2002).
In this sense our simulation is only a lower limit to
the arc frequency. We shall investigate this
effect quantitatively in a forthcoming paper 
(Bode, Ostriker, \& Wambsganss, in preparation).

\section{Discussion }

In B98 the observational
situation concerning the frequency of giant
arcs is described and discussed in detail, with particular respect
to the EMSS survey  (Gioia \& Luppino 1994, Le Fevre et al. 1994).
In a new  observational study, 
Zaritsky \& Gonzalez (2003) have  recently 
analyzed the occurrence of arcs around clusters
in the Las Campanas Distant Cluster Survey.
They found at least two giant arcs with a length-to-width
ratio of 
$r_{\mathrm{arc}} \ge 10$
in the sample
of clusters within a redshift range $0.5 \le z_{\mathrm{cl}} \le 0.7$,
plus one strong lensing system outside this range.
Their result confirms and strengthens the previous result
that the observed frequency of arcs exceeds the B98
predictions by a factor of about 10.
They also discuss in detail previous lensing surveys, 
which consistently point in the same direction.
Gladders et al. (2003) likewise confirm that the incidence
of arcs found
in the Red-Sequence Cluster Survey ``shows significant disagreement
with theoretical predictions''.

With regard to this disagreement,
the results of the numerical investigation
presented  in the previous section
on the frequency of gravitationally lensed
arcs with given length-to-width ratios in an LCDM
cosmology have three important aspects:

\begin{enumerate}

\item We find that the {\bf optical depth} for the occurrence
	of giant arcs with a 
	length-to-width ratio
	of at least $r_{\mathrm{arc}} \ge 10$ 
	{\bf for galaxies/sources at a redshift of $\mathbf{z_s=1.0}$ is} 
	slightly higher, but 
	{\bf essentially the same as
	what B98 have found},
	when corrected for the slight difference in the adopted
	cosmological model.

\item
	{\bf The lensing optical depth is 
	a {\em very steep} function of the source redshift
	} for $0.5 \le z_s \le 2.5$, 
	and it is increasing even further for $z_s > 2.5$.
\item
	Combining these two aspects with a 
	realistic galaxy/source redshift 
	distribution that extends beyond 
	$z_s=1.0$ (and
	allows for a fraction of the galaxies to be at
	$z_s \ge 1.5$)  naturally results in a much 
	higher probability for giant arcs than previously
	proposed for LCDM models.

\end{enumerate}
The second aspect above is the key to the solution of the puzzle,
and the reason that our conclusions with regard to the 
frequency of arcs in an LCDM cosmology
are different from those of B98.
They argue that the 
exact source redshift has only very little influence
on the resulting optical depth for long arcs. 
The average cluster redshift in their models is
$z_{\mathrm{cluster}} \approx 0.3 \ \mathrm{ to } \ 0.4$, and
they point out that the critical surface mass density 
$\Sigma_{\mathrm{crit}}$ 
for multiple images changes only little 
when sources are shifted from $z_s=0.8$ to $z_s=1.2$.
In fact, for a lens redshift of e.g. $z_L=0.4$,
the critical surface mass density 
does indeed change by only about 25\% 
between these two source redshift values.
However, 
the most important aspect here is not the fractional decrease of 
the critical surface mass density $\Sigma_{\mathrm{crit}}$. 
What matters is rather 
how many more galaxy clusters become super-critical when
the critical surface mass density is lowered by that amount. 
The cluster mass function at the high mass end 
is {\em very} steep, 
with a logarithmic slope of about -5 or even steeper
(see, e.g., Bahcall, Fan \& Cen 1997, or Jenkins et al. 2001).
This means that even a relatively modest decrease in
the value of the critical surface mass density for
source redshifts
slightly larger than unity can result
in a relatively large increase of clusters with 
central surface mass density above this value, 
hence producing strongly lensed arcs
(cf. Li and Ostriker 2002).
The resulting values for the optical depth of
arcs with different
$\mu_{\mathrm{arc}}$ are
shown in Figure 1 for various source redshifts; the
corresponding numerical values are listed in Table 1.
In particular, we find that for a source redshift of $z_s=1.5$
the optical depth for arcs (with length-to-width ratio
$r_\mathrm{arc} \ge 10$) is increased by about 
a factor of six, compared to its value for $z_s=1.0$!

Another factor that contributes slightly to an increase of the 
probability of arcs for higher source redshifts
is that the extra matter added at
higher redshifts can boost non-critical clusters
over the threshold for multiple images and/or strongly
magnified arcs. This effect---
which is automatically included by our method---
can be surprisingly large  for
higher source redshifts
(Wambsganss, Bode \& Ostriker, in preparation).

It is important to note that
the redshifts of known luminous arcs extend
out to redshifts beyond $z_{\mathrm{arc}} \ge 5$.  
Table 2 lists a selection of clusters with 
measured arc redshifts.
This is only meant to serve
as an illustrative example, since most of these lens systems
were not found in any systematic arc searches.
But, even though we cannot take the redshift distribution
of these arcs as representative, 
the fact that there
are so many arcs with redshifts beyond 
$z_\mathrm{arc} \ge 1.5$ and extending
to $z_\mathrm{arc} \approx 5$,
clearly  means that high redshift sources have
to be included in the determination of the 
overall optical depth for arcs.
The very recent  result of 
Gladders et al. (2003) confirms 
this trend of high arc redshifts:  
their arcs with measured redshifts 
range between
$1.7 \le z_{\rm arc} \le 4.9$, and
the correspondingly  very large
distances of the
RCS lensing clusters for which no
arc redshift could be measured yet 
($0.6 \le z_{\rm cluster} \le 1.2$)
point to (very) high redshifts
for those arcs, too.

To explore the effect of including higher redshift sources
more quantitatively,
we defined  three ``test cases'' for different redshift distributions
of sources, convolved them with the corresponding
optical depths, and compared the resulting value of
the total optical depth with $\tau_{\mathrm{B98}}$,
the value obtained by B98. 
The three cases are: (1) sources divided evenly between three
redshifts $z_{\mathrm{arc}}$=0.5, 1.0, and 1.5, with a
resulting optical depth 
of $\tau_{\mathrm{case 1}}=9.3 \times 10^{-7}\approx 
2.8\times \tau_{\mathrm{B98}}$;
(2)  sources divided evenly between three
redshifts $z_{\mathrm{arc}}$=1.0, 1.5, and 2.5, with a
resulting optical depth 
of $\tau_{\mathrm{case 2}}= 3.3\times 10^{-6}\approx 
10\times \tau_{\mathrm{B98}}$;
and (3)
33\% of sources at $z_{\mathrm{arc}}=0.5$, 
30\% at $z_{\mathrm{arc}}=1.0$, 
20\% at $z_{\mathrm{arc}}=1.5$, 
10\% at $z_{\mathrm{arc}}=2.5$, 
 5\% at $z_{\mathrm{arc}}=3.7$,  and
 2\% at $z_{\mathrm{arc}}=4.8$, which yields an
optical depth of $\tau_{\mathrm{case 3}}= 2.4\times 10^{-6}\approx 
7.3\times \tau_{\mathrm{B98}}$. 
Our case 1 still has an average value
of $<z_{\mathrm{arc}}>=1.0$, but
due to the ``spread'' in redshift, the value of the
optical depth is increased by almost a factor of three 
compared to the case with ``all arcs at redshift one''.
The other two cases, which allow for a fraction
of arcs with 
higher redshift, produce increases of factors of 10 and 7.3,
when compared to the 
B98 values of the optical
depth.   

The examples of the previous paragraph show that
for a realistic 
source redshift distribution which extends out to 
$z_{\mathrm{arc}} \ge 1.5$,  we get about an order of
magnitude more arcs for a LCDM concordance model. 
In particular
for our test cases 2 and 3, the higher optical
depth results in between 2000 and 3000 
strongly lensing cluster producing arcs
across the sky, or one arc per 14 to 20 square degrees.
This is perfectly consistent with the observational studies
of Gioia \& Luppino (1994), Gladders et al. (2003) and 
Zaritsky \& Gonzalez (2003).
Gladders et al. (2003) had realized that their sample ``probes
a somewhat different redshift range'' compared to the assumptions
of B98. They had suggested that ``the resulting differences
should be factors of order unity''. 
We show here that 
relaxing the assumption that
all the sources have a redshift of $z_s=1$, i.e.
allowing for a realistic source population 
with redshifts well beyond unity, means that  
an LCDM model
can easily account for a factor of ten more arcs than
was previously predicted.

\section{Summary}

We performed ray-shooting simulations through a 
pseudo-3D matter distribution  obtained for 
a concordance lambda-dominated flat cosmological model (LCDM).
We determined the frequency of giant arcs,
as a tool to test the predictions of a 
such an LCDM cosmology. 
When we put all our sources at a redshift of
$z_s=1.0$, we basically reproduce the results of Bartelmann et al. (1998)
for the frequency of arcs with a 
length-to-width ratio of at least ten:
our results are about 12 \% higher than theirs, which can be
understood as a consequence 
of a slightly higher normalization
and the inclusion of the full 3D matter distribution.

However, we find that the 
{\em optical depth is a 
very steep function of the source redshift}. 
The contribution  of
sources at redshifts beyond one easily dominates the statistics. 
Even for an {\em average} redshift of the lensed galaxies
of roughly $<z_s> \approx 1.0$ with a broad distribution,
the frequency of arcs could be increased easily by
a factor of three.

The redshift distribution of the known luminous arcs shows
that many of them {\em are} at $z_s \ge 1.5$.
If we hence give up the constraint that all viable sources
for giant arcs are at redshift one,
and allow for a distribution
of source redshifts, 
then we find good  agreement between the 
frequency of arcs predicted by an LCDM cosmological model
and the observed situation, as evaluated by Gioia \& Luppino 
(1994),  Zaritsky \& Gonzalez (2003) or Gladders et al. (2003).
The predicted frequency of giant arcs for
a concordance LCDM model goes up by about an order of magnitude
compared to previous estimates (Bartelmann et al. 1998)
and results in about one arc per 20 square degrees.
Hence we consider the ``missing arc''  problem for
a concordance  LCDM cosmology to be solved.

\acknowledgments

We are pleased to acknowledge useful discussions with 
Neta A. Bahcall, 
Renuye Cen,
Bohdan Paczy\'nski, and
Massimo Meneghetti.
This research was supported by the National Computational Science
Alliance under NSF Cooperative Agreement ASC97-40300, PACI Subaward 766;
also by NASA/GSFC (NAG5-9284).  Computer time was provided by NCSA
and the Pittsburgh Supercomputing Center.


\begin{figure}
\plotone{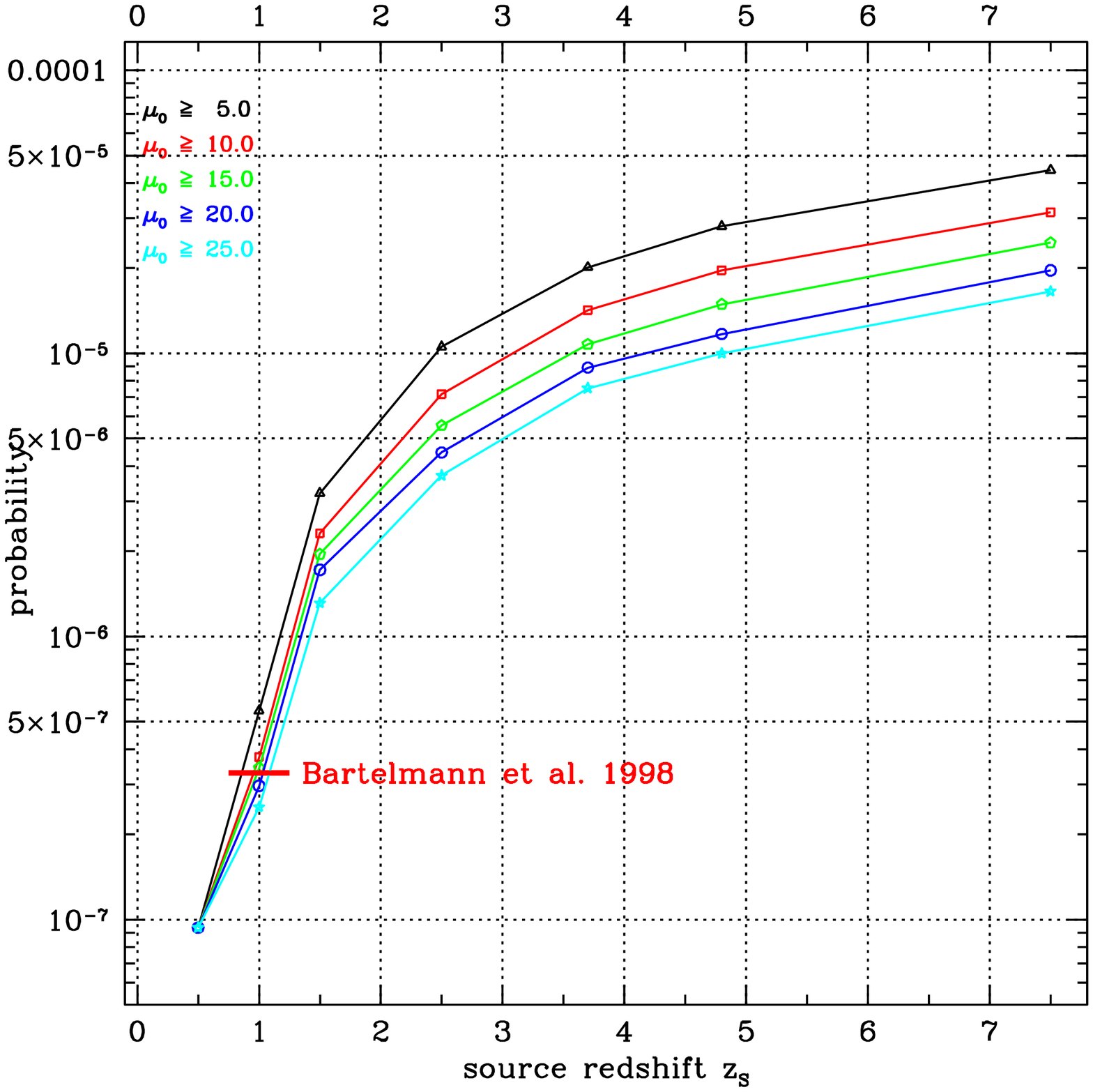}
\caption{Probability for the occurrence of gravitationally
lensed arcs with length-to-width ratios $r$ 
($\approx$ magnifications $\mu_i$) of 
$\ge 5$ (triangles, black), 
$\ge 10$  (squares, red), 
$\ge 15$  (pentagons, green),
$\ge 20$  (circles, blue), and 
$\ge 25$  (stars, cyan) 
	as a function of galaxy (source) redshift, for the LCDM concordance
	model. For comparison, the probability as determined
	by Bartelmann et al. (1998) is indicated
	as a red horizontal bar; 
	they had assumed that all galaxies are at redshift $z_s=1$ and
	evaluated arcs with length-to-width ratio $ r \ge 10$.}
\end{figure}

\clearpage

\begin{deluxetable}{ccclrrrrrrrr}
\tabletypesize{\scriptsize}
\tablecaption{Optical depth for arcs as a function of source
redshift in a  concordance 
LCDM model: \label{tbl-1}}
\tablewidth{0pt}
\tablehead{ \colhead{length-to-width ratio\tablenotemark{a}} 
& \colhead{source redshift: }   \\ 
& \colhead{$z_s = 0.5$}   
& \colhead{1.0 } &\colhead{1.5}   
& \colhead{2.5 } &\colhead{3.7}   
& \colhead{4.8 } &\colhead{7.5}   
}
\startdata

$  r_{\mathrm{arc}} \ge  5$ & 9.4$\times 10^{-8}$ & 5.5$\times 10^{-7}$ & 3.2$\times 10^{-6}$ & 1.1$\times 10^{-5}$ & 2.0$\times 10^{-5}$ & 2.8$\times 10^{-5}$ & 4.4$\times 10^{-5}$ \\
$  r_{\mathrm{arc}} \ge 10$ & 9.4$\times 10^{-8}$ & 3.8$\times 10^{-7}$ & 2.3$\times 10^{-6}$ & 7.2$\times 10^{-6}$ & 1.4$\times 10^{-5}$ & 2.0$\times 10^{-5}$ & 3.2$\times 10^{-5}$ \\
$  r_{\mathrm{arc}} \ge 15$ & 9.4$\times 10^{-8}$ & 3.5$\times 10^{-7}$ & 2.0$\times 10^{-6}$ & 5.6$\times 10^{-6}$ & 1.1$\times 10^{-5}$ & 1.5$\times 10^{-5}$ & 2.5$\times 10^{-5}$ \\
$  r_{\mathrm{arc}} \ge 20$ & 9.4$\times 10^{-8}$ & 3.0$\times 10^{-7}$ & 1.7$\times 10^{-6}$ & 4.5$\times 10^{-6}$ & 8.9$\times 10^{-6}$ & 1.2$\times 10^{-5}$ & 2.0$\times 10^{-5}$ \\
$  r_{\mathrm{arc}} \ge 25$ & 9.4$\times 10^{-8}$ & 2.5$\times 10^{-7}$ & 1.3$\times 10^{-6}$ & 3.7$\times 10^{-6}$ & 7.5$\times 10^{-6}$ & 1.0$\times 10^{-5}$ & 1.7$\times 10^{-5}$ \\

 \enddata

\tablenotetext{a}{We assume here that the 
length-to-width ratio of the arc 
is equal to its magnification}

\end{deluxetable}

\begin{deluxetable}{lcclrrrrrrrr}
\tabletypesize{\scriptsize}
\tablecaption{Redshift information on known arcs\tablenotemark{a}: \label{tbl-2}}
\tablewidth{0pt}
\tablehead{ \colhead{Cluster} & \colhead{Cluster redshift}   & \colhead{Arc  redshift } &\colhead{reference}   } 
\startdata 
A370      & z$_{\mathrm{cl}}=0.374$ & z$_{\mathrm{arc}}$ = 0.735, 0.81, 1.3 & Soucail et al. (1988), B{\'e}zecourt et al. (1999)\\
A963      & z$_{\mathrm{cl}}$ = 0.206 & z$_{\mathrm{arc}}$ = 0.771 & Ellis et al. (1991) \\ 
A2163     & z$_{\mathrm{cl}} = 0.201 $ & z$_{\mathrm{arc}}$ = 0.728 & Miralda-Escude \& Babul (1995)\\ 
A2218     & z$_{\mathrm{cl}} = 0.17 $ & z$_{\mathrm{arc}}$ = 0.70, 1.03, 2.52, 5.60& Pell{\'o} et al. (1992), Ebbels et al. (1996), Natarajan et al. (2002) \\
A2390     & z$_{\mathrm{cl}} = 0.23$ & z$_{\mathrm{arc}}$ = 0.913, 4.04 & Frye \& Broadhurst (1998), Pell{\'o} et al. (1999) \\
AC114     & z$_{\mathrm{cl}}=0.312$ & z$_{\mathrm{arc}}$ = 1.69, 1.87, 3.347& Pell{\'o} et al. (2001), Campusano et al. (2001) \\ 
CL0024+17 & z$_{\mathrm{cl}}$ = 0.39 & z$_{\mathrm{arc}}$ = 1.675 & Broadhurst et al. (2000) \\
CL0500-24 & z$_{\mathrm{cl}}$ = 0.327 & z$_{\mathrm{arc}}$ = 0.913 & Schindler \& Wambsganss (1997) \\
CL1358+62 & z$_{\mathrm{cl}}$ = 0.328& z$_{\mathrm{arc}}$ = 4.92 & Franx et al. (1997) \\
CL2236-04 & z$_{\mathrm{cl}}$ = 0.56 & z$_{\mathrm{arc}}$ = 1.12, 1.33 & Kneib et al. (1994) \\
CL2244-02 & z$_{\mathrm{cl}} = 0.328$ & z$_{\mathrm{arc}}$ = 2.24 & Mellier et al. (1999) \\ 
MS0440+0204 & z$_{\mathrm{cl}} = 0.190$ & z$_{\mathrm{arc}}$ = 0.532, 1.63 & Gioia et al. (1998), Luppino et al. (1999) \\
MS1512+36 & z$_{\mathrm{cl}} = 0.37$ & z$_{\mathrm{arc}}$ = 2.72 & Seitz et al. (1998) \\
MS2137-23 & z$_{\mathrm{cl}} = 0.313$ & z$_{\mathrm{arc}}$ = 1.50 & Sand  et al. (2002) \\
PKS0745-191 & z$_{\mathrm{cl}}= 0.103 $ & z$_{\mathrm{arc}}$ = 0.433& Allen et al. (1996)\\
RCS0224-002 & z$_{\mathrm{cl}}= 0.773 $ & z$_{\mathrm{arc}}$ = 1.7, 4.88 & Gladders, Yee \& Ellingson (2002), Gladders et al. (2003) \\
RCS2319-004 & z$_{\mathrm{cl}} \approx 1.0 $ & z$_{\mathrm{arc}} \approx 3-4$ & Gladders et al. (2003) \\
RXJ1347-1145& z$_{\mathrm{cl}}= 0.451 $ & z$_{\mathrm{arc}}$ = 0.81 & Sahu et al. (1998); Ravindranath \& Ho (2002) \\
1E0657-56 & z$_{\mathrm{cl}} = 0.296 $  & z$_{\mathrm{arc}}$ = 2.34, 3.08, 3.24 & Mehlert et al. (2001) \\
 \enddata

\tablenotetext{a}{This list is meant to illustrate the
broad redshift distribution of the known arcs; it is not complete 
in any sense}

\end{deluxetable}

\end{document}